\documentclass[preprint,showpacs,preprintnumbers,amsmath,amssymb]{revtex4}

\usepackage{graphicx}
\usepackage{dcolumn}
\usepackage{bm}

\begin{document}

\preprint{}

\title{Two dimensional curved disks on a sphere: the evolution of kinetic energy}

\author{ J.R.Darias$^{1,2}$ and  N.Olivi-Tran$^{1,3}$} 
\affiliation{$1$Laboratoire de Physique de la Mati\`ere Condens\'ee et Nanostructures, UMR-CNRS 5586, Universit\'e Claude Bernard Lyon I, Domaine Scientifique de la Doua, 69622 Villeurbanne cedex, France \\ 
$2$Departamento de F\'isica, Universidad Sim\'on Bol\'ivar, Apartado 89000, Caracas 1080-A, Venezuela \\
$^3$Laboratoire de Sciences des Proc\'ed\'es C\'eramiques et Traitements de Surface, UMR-CNRS 6638, Ecole Nationale 
Sup\'erieure de C\'eramiques Industrielles, 47 avenue Albert Thomas, 87065 Limoges cedex,
France }

\date{today}

\begin{abstract}
We put disks on a sphere between two parallels of this sphere. The disks have the curvature of the sphere
and interact via a simplified Hertz law. We analyze the behavior of the total kinetic energy of the whole
assembly of disks with Riemanian geometry and contrarily to flat spaces, this kinetic energy increases. This is due to a non radial
component of the resulting velocity vector at each interaction between two disks.
The study of granular matter in twodimensional curved spaces is therefore not trivial.

\end{abstract}

\vfill
\maketitle

Granular matter has been a wide subject of interest since the 80's \cite{kada,nagel,jaeger}.
But up to now no study of granular matter and in particular soft particles has been
made in a space with curvature.
The rare studies which deal with curved spaces are the articles of Caillol \cite{caillol}
for plasmas on spheres and hyperspheres. Although, these studies are very different
from granular matter analysis as electrons have long range interactions. Therefore
the fact that the space are spheres or hyperspheres is linked to boundary conditions \cite{caillol2}.

For granular matter, the interactions are short ranged and even if one tried to compare periodic
boundary conditions for a plane with a sphere, the curvature interferes and the comparison
can not be direct (essentially because of the metric).

Here we used a Molecular Dynamics program for soft disks on a sphere. A simplified Hertz law has been
used in order to match the metric of differential geometry of the sphere.
In order to check that our results were not an artifact of the computational program
we used two algorithm for the integration: the predictor corrector and the velocity Verlet algorithms.
Both gave the same results.

We put $N$ disks with equal radius on spheres with different radii. The spheres match exactly the curvature of the sphere,
i.e. the space where they are enclosed is the sphere.
The number of spheres may vary. The surface of the spheres where they can move is limited by two parallels
of the spheres at equal distance from the equator, and that in order two avoid the poles which are critical points
for the calculations (see fig.1). Gravity acts from north pole on the top of the sphere
to south pole on the bottom of the sphere.
The disks are soft and interact via a simplified Hertz law in a curved space.

In order to take account of the curvature, we used the following metric tensor of Riemanian geometry:
\begin{eqnarray}
T_{11}=1 \\ T_{22}=r^2 \\ T_{33}=r^2 \sin^2 \theta
\end{eqnarray}
The other components of the tensor are equal to zero.
For all physical parameters included in the following equations and using lengths dimensions,
we used this tensor.

The simplified Hertz law is written also as follows, always in the spirit of the metric tensor:
\begin{equation}
{\bf F_{ij}}=[K_n(d-r_{ij})-m_e \gamma_n ({\bf v_{ij}.n})]{\bf n} \\
+ \epsilon min[\gamma_s |{\bf v_{ij}.s}|, \mu |{|\bf F_n}|]{\bf s}
\end{equation}
with ${\bf F_{ij}}$ is the relative force between disks $i$ and $j$, $K_n$ is a spring constant,
$d$ is the diameter of the disks, $r_{ij}$ is the center to center distance between disks $i$ and $j$,
$m_e$ is the mass of one disk, $\gamma_n$ (resp. $\gamma_s$) is the dissipation coefficient in the normal ( resp. shear)direction, ${\bf n}$ (resp. ${\bf s}$)
is the normal (resp. shear) vector (with respect to two disks), and ${\bf F_n}$ is the normal force.
\begin{equation}
\epsilon=-\frac{{\bf v_{ij}.s}}{|{\bf v_{ij}.s}|}
\end{equation}
$\epsilon$ is the sign of the relative velocity in the shear direction ${\bf s}$.

The two parallels are located at $\theta = \pi/2$ and $\theta = 3 \pi/2$.
The disks interact with the parallels with a damping coefficient equal to 0.75.
The data are have no dimensions taking account of their mass $m$, the gravity $g$ and
their diameter $d$.
So, $K_n=2000mg/d$, $\gamma_n=25 (d/g)^{1/2}$,$\gamma_s=25 (d/g)^{1/2}$ and $\mu=0.45$.

The normal vector is calculated as follows:
\begin{equation}
{\bf n}={\bf d \ell}= \frac{\vec{d \ell}}{|\vec{d \ell}|}=\frac{r d \theta}{|\vec{d \ell}|} {\bf \theta}
+\frac{r \sin \theta d \phi}{|\vec{d \ell}|} {\bf \phi}
\end{equation}
and the shear vector is calculated as follows:
\begin{equation}
{\bf s}=\frac{r \sin \theta d \phi}{|\vec{d \ell}|} {\bf \theta} - \frac{r d \theta}{|\vec{d \ell}|} {\bf \phi}
\end{equation}
therefore ${\bf n.s}=0$ in spherical coordinates $r,\theta$ and $\phi$.

At the beginning of computation, i.e. at Molecular Dynamics Step (MDS) equal to 1, the disks are put at random 
between the two parallels with a random distribution of velocity vectors in all directions.
Gravity is acting since the beginning of computation.

When the disks are put at random within the two parallels and do not see each other, we checked that they fall
to the bottom parallel until reaching it.
We also checked that two disks which have a trajectory exactly on the equator with the same velocity vector
norm but with inverse direction, the acceleration of the two disks did fall to zero as well as their kinetic
energy.
With the same program, but with a radius of the sphere infinite (a plane), we checked that the kinetic
energy deduced from the acceleration of the disks was tending to zero as is usually find in a plane.

The results are shown in figures 2 and 3.
Figure 2 corresponds to different radii of the sphere but with a constant number of disks.
As one can see, for increasing radius of the sphere, the maximum of the kinetic energy
is lower than for a smaller radius of the sphere.
Figure 3 corresponds to different number of disks but with an equal radius of the sphere.
Here again, for an increasing number of disks the resulting maximum kinetic energy is larger.

Let us have an explanation of this phenomenon of increasing kinetic energy (not tending to zero
though the existence of dissipation in the Hertz law).
For that, we plotted figure 4 which represents two disks in interaction on the surface of the sphere
but with different norms of the velocity vector after interaction.
If one calculates the resulting vector of the two interacting velocity vectors after interaction,
one can see that this resulting vector has a non radial component with respect to the center
of the sphere.
This means that at after each interaction, the non radial component obtained will give place
to a tangential component of the resulting velocity vector of each disk. This fact will give
an additional term to the total kinetic energy of the whole assembly of disks.
Therefore the kinetic energy will increase infinitely. In our case, as we use the Hertz law
with dissipation, the kinetic energy will stabilize at a value where the number of interactions
and the value of the additive kinetic energy for each time step will be counterbalanced
by the dissipation.

In this letter, we showed that unlike plane spaces, in a curved space which is a sphere,
the total kinetic energy of a whole assembly of disks which are on the sphere and have the
curvature of the sphere increases until reaching a stable value in time. This value is different
from zero and is higher for larger numbers of disks and smaller sphere radii.
The explanation of this phenomenon is the non radial component which appears at each
interaction following the Hertz law, between two disks, and which leads to an additive
non zero kinetic energy to the total kinetic energy.
Studying twodimensional disks in a space with non zero curvature is therefore not trivial
and very different from the study of granular matter in flat spaces.

Acknowledgement 
We would like to acknowledge financial support for the PCP {\it Statique et
dynamique des milieux granulaires}

\pagebreak

\pagebreak

\begin{figure}
\centering
\resizebox*{14cm}{14cm}{
\includegraphics[width=10cm]{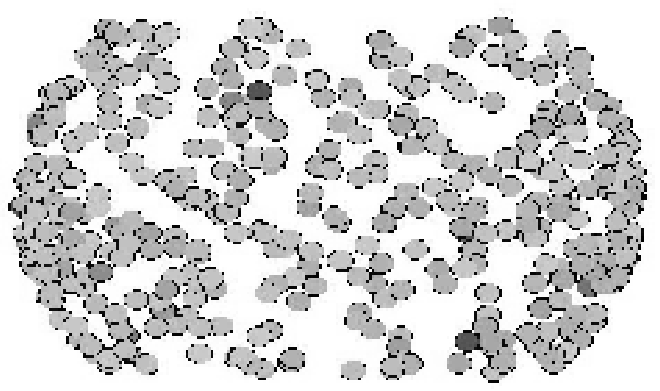}}
\caption{Schematic representation of the disks on the sphere limited by two parallels. The disks are represented
as small spheres because of software displaying
 }
\end{figure}

\begin{figure}
\includegraphics[width=8cm]{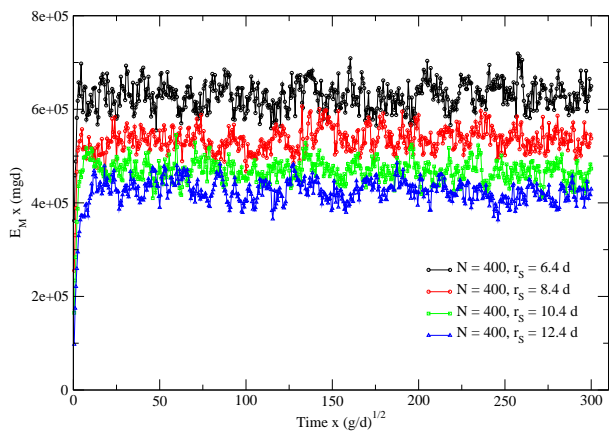}
\caption{Total kinetic energy as a function of time for the different sphere radii (in unit of disk diameter) but a constant number of disks}
\end{figure}

\pagebreak

\begin{figure}
\includegraphics[width=8cm]{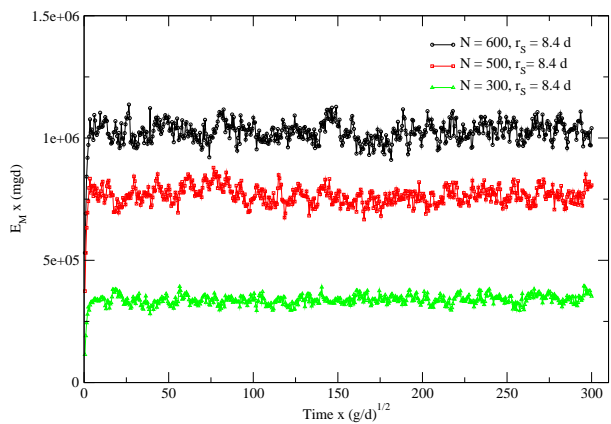}
\caption{Total kinetic energy as a function of time for a constant sphere radius (in unit of disk diameter) but different number of disks}
\end{figure}

\pagebreak

\begin{figure}
\includegraphics[width=10cm]{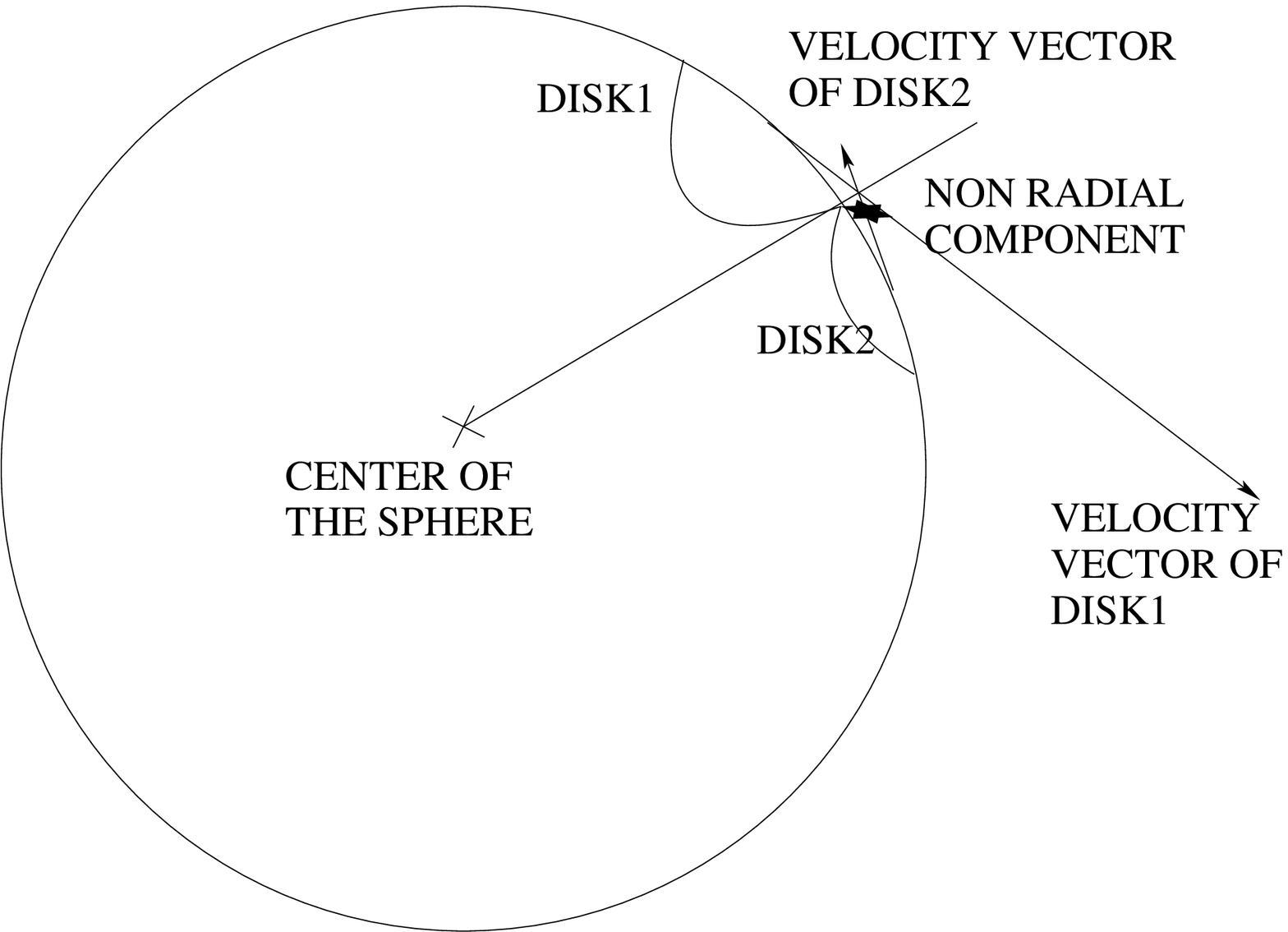}
\caption{Schematic representation of the non radial component of the relative velocity vector
after an interaction between two disks on the sphere}
\end{figure}


\begin{thebibliography}{99}
\bibitem{kada}{L. P. Kadanoff, Review of Modern Physics, {\bf 71}, 435, (1999); and references within}
\bibitem{nagel}{S. Nagel, Review of Modern Physics, {\bf 64}, 321, (1992); and references within}
\bibitem{jaeger}{H. M. Jeager, S. R. Nagel and R. P. Behringer, Physics Today, {\bf 49}, 32, (1996); Review of Moderm Physics, {\bf 68}, 1259, (1996); and references within}
\bibitem{caillol}{J. M. Caillol, J. Chem. Phys. {\bf 111} 6528 (1999)}
\bibitem{caillol2}{J. M. Caillol and D.Gilles, J. of Stat. Phys. {\bf 100} 905 (2000)}

\end{thebibliography}
\end{document}